\documentclass[conference]{IEEEtran}
\IEEEoverridecommandlockouts

\usepackage{cite}
\usepackage{amsmath,amssymb,amsfonts}
\usepackage{algorithmic}
\usepackage{graphicx}
\usepackage[caption=false,font=footnotesize]{subfig}
\usepackage{textcomp}
\usepackage{multirow}
\usepackage{xcolor}
\usepackage{colortbl}

\def\BibTeX{{\rm B\kern-.05em{\sc i\kern-.025em b}\kern-.08em
    T\kern-.1667em\lower.7ex\hbox{E}\kern-.125emX}}

\begin{document}

\title{FR3 for 6G Networks: A Comparative Study against FR1 and FR2 Across Diverse Environments}

\author{
Fahimeh Aghaei$^{*}$, Mehdi Monemi$^{*}$, Mehdi Rasti$^{*}$, Murat Uysal$^{\dagger,\ddagger}$\\[1ex]
\small $^{*}$\textit{Centre for Wireless Communications (CWC), University of Oulu, Oulu, Finland}\\
\small $^{\dagger}$\textit{Division of Engineering, New York University Abu Dhabi (NYUAD), Abu Dhabi 129188, UAE}\\
\small $^{\ddagger}$\textit{Research Institute NYUAD Wireless Center, Abu Dhabi 129188, UAE}\\
\small fahimeh.aghaei@oulu.fi, mehdi.monemi@oulu.fi, mehdi.rasti@oulu.fi, murat.uysal@nyu.edu
\thanks{This paper is supported by the Finnish Ministry of Education and Culture through the Intelligent Work Machines Doctoral Education Pilot Program (IWM VN/3137/2024-OKM-4); the Research Council of Finland (former Academy of Finland) via 6G Flagship (Grant No. 369116) and Profi6 Project (Grant No. 336449); and the Business Finland via the 6GBridge - Local 6G project (Grant No. 8002/31/2022) and DigiPave project with (Grant No. 3992/31/2023). The work of M. Uysal is supported by Tamkeen under the Research Institute NYUAD grant CG017.}
}

\maketitle

\begin{abstract}
Motivated by increasing wireless capacity demands and 6G advancements, the newly defined Frequency Range 3 (FR3, 7.125-24.25 GHz), also known as the upper mid-band, has emerged as a promising spectrum candidate. It offers a balance between the large bandwidth potential of millimeter-wave bands and the favorable propagation characteristics of sub-6 GHz bands. As a result, the upper mid-band presents a strong opportunity to enhance both coverage and capacity, particularly for 6G systems and Cellular Vehicle-to-Base Station (C-V2B) communications. Harnessing this potential, however, requires addressing key technical challenges through accurate and realistic channel modeling across diverse urban environments, including Suburban, Urban, and HighRise Urban scenarios.
To this end, we employ a ray-tracing tool to characterize downlink propagation and enable detailed channel modeling for reliable C-V2B links. We evaluate data rate performance across FR1 (sub-6 GHz), FR3, and FR2 (mmWave) bands using antenna array configurations designed for different urban environments. The results show that, under equal aperture sizes, FR3 achieves higher data rates than FR2 for cell-edge User Equipment (UEs) in both interference-free and full-interference scenarios, indicating that the additional array gain at mmWave is insufficient to fully compensate for the severe experienced path loss. Integrating one-hand-grip pedestrian UEs model into ray tracer shows that transitioning from vehicular to pedestrian UEs results in negligible differences in coverage probability (about 1\%--3\%) across all frequencies, with the minimum differences observed in FR3, particularly at 8.2 GHz.
\end{abstract}


\section{Introduction}

As sixth-generation (6G) networks advance to support ultra-reliable and high-capacity services in dynamic environments, the demand for radio spectrum has reached a critical point. Driven by data-intensive applications such as fixed wireless access (FWA) and metaverse services which require substantially higher data rates than conventional mobile usage \cite{Ref1}, \cite{Ref2}, the search for viable 6G spectrum has increasingly focused on the upper mid-band (7.125--24.25 GHz), referred to as Frequency Range 3 (FR3) \cite{RefFR3}. This spectrum range represents a viable candidate for 6G spectrum allocation, balancing multi-gigahertz bandwidth for next-generation throughput with sufficiently robust propagation to support reliable connectivity in dense, interference-limited urban environments \cite{SurveyFr3}.

Urban 6G performance analysis centers on evaluating Key Performance Indicators (KPIs), such as signal quality and coverage across diverse city environments characterized by varying building densities and architectural heights \cite{Ref23}. In these settings, the interaction between radio waves and the complex environment including buildings, skyscrapers, trees, and other reflective or absorptive surfaces intensifies multifaceted propagation phenomena such as high-order multipath scattering, diffraction, and material-specific attenuation. These static structures, combined with dynamic obstacles like vehicles and pedestrians, trigger frequent transitions between Line-of-Sight (LoS) and Non-Line-of-Sight (NLoS) states. Consequently, an assessment of 6G network reliability necessitates precise, frequency-dependent channel modeling that effectively captures how these environmental and temporal complexities shape the propagation environment across various bands.

Accurate channel modeling is fundamental to the design and performance evaluation of next-generation wireless systems across different urban environments. The 3rd Generation Partnership Project (3GPP) TR 38.901 report provides a standardized framework for stochastic modeling across sub-6 GHz, mmWave, and upper mid-band frequencies, effectively capturing critical propagation effects such as shadowing and delay spread \cite{Ref3}. Although these models are widely used for large-scale coverage analysis, they frequently lack the spatial and temporal resolution required to evaluate advanced 6G features such as massive Multiple-Input Multiple-Output (MIMO) and adaptive beamforming in dynamic vehicular contexts \cite{monemi2025higher,RefNarges}. Ray tracing (RT) methods address these limitations by offering a site-specific deterministic approach that simulates electromagnetic interactions with the actual urban geometry. This fine-grained level of detail is particularly vital for analyzing dynamic microcell scenarios where physical obstructions and moving obstacles fundamentally alter the channel state.

To advance 6G development, recent research has examined wireless channel characteristics across a variety of frequency bands and deployment scenarios. For example, \cite{Refmatlab} utilized RT to analyze 2.4 GHz air-to-ground channels for aerial base stations (BSs), identifying that path loss exponents and shadow fading are heavily dependent on the environment (e.g., Suburban vs HighRise Urban). Moving to higher frequencies, \cite{RappFr3} performed extensive measurements in urban microcells at 6.75 GHz (FR1C) and 16.95 GHz (FR3). Their analysis indicates that the upper mid-band exhibits lower path loss exponents than mmWave or sub-THz frequencies, while delay and angular spreads tend to decrease as frequency increases. These findings support the argument by \cite{MainRef} that the upper mid-band offers a critical trade-off between coverage and capacity, necessitating precise channel models for realistic system design.

In this paper, we present a comprehensive FR3 channel characterization across three environments including HighRise Urban, Urban, and Suburban environments compared with sub-6 GHz and mmWave bands. Using RT techniques, we focuse on Cellular Vehicle-to-Base Station (C-V2B) downlink communication scenario, where BSs transmit to vehicles acting as User Equipments (UEs)\footnote{Considering UEs with identical antenna gains, the performance in vehicular networks typically falls between that of fixed-mounted devices (e.g., fixed sensors) which benefit from stable placement and the absence of metallic body of the vehicle, and that of handheld UEs (e.g., pedestrians holding cellphones), which experience higher signal variability due to hand grip attenuation.}.
The results then compare to the coverage performance in Cellular Pedestrian-to-Base Station (C-P2B) downlink communication scenario, where BSs serve pedestrian as UEs.
Motivated by \cite{Refmyicc, Ref9} employing RT for mmWave and vehicular studies, we adopt this approach with detailed 3D CAD models of vehicles, humans, and the different urban layouts. MIMO technology is integrated to assess beamforming impacts on signal quality, data rate, and coverage. Simulations are conducted using Remcom’s Wireless InSite \cite{Ref6} to evaluate signal-to-noise ratio (SNR), and signal-to-interference-plus-noise ratio (SINR), subsequently deriving data rate ($R$) under different interference scenarios. The main contributions of this work are:

\begin{itemize}
    \item To facilitate the integration of FR3 into 6G networks, we perform a comprehensive C-V2B achievable data rate analysis across the FR3 spectrum (8.2 GHz and 15 GHz), benchmarked against sub-6 GHz (4.6 GHz) and mmWave (28 GHz) bands using a RT tool. The study considers both interference-free and full-interference cases employing a static blockage model resulted from buildings based on the statistical ITU model across Suburban, Urban, and HighRise Urban environments. A particular focus is placed on HighRise Urban cases, where a realistic 3D CAD model of Dubai downtown is integrated and validated against the corresponding statistical ITU model. To ensure a fair comparison, we assume an equal aperture size across all frequencies. {\it The comparison of the data rate across HighRise Urban area between the sub-6 GHz, FR3, and mmWave shows that for {\bf cell-edge} UEs, where the average data rate is typically the lowest, FR3 provides a higher performance. Therefore, although mmWave offers a higher array gain due to the incorporation of larger number antenna elements, this gain cannot fully offset the severe path loss at corresponding high frequencies.}
    \item The integration of an one-hand-grip pedestrian UE model aligned with 3GPP TR 38.901 into the urban CAD environment reveals that the transition from vehicular to pedestrian UEs yields no significant changes in coverage performance. Our findings indicate that coverage probability is relatively insensitive to UE mobility and height, with a negligible differences of approximately 1\% to 3\% across the evaluated frequencies. Notably, this deviation is minimized in the FR3 band, particularly at 8.2 GHz, suggesting that FR3 exhibits greater resilience to variations in user height and mobility compared to FR1 and FR2.
\end{itemize}

The organization of the paper is as follows: Section II describes the city layout and scenarios, followed by the channel modeling methodology in Section III. Statistical evaluation of downlink C-V2B and C-P2B performance is presented in Section IV, and Section V offers concluding remarks.

\section{City Layout and Scenarios}

Accurate RF modeling in urban environments requires a detailed representation of building topologies. According to the International Telecommunication Union (ITU-R) recommendation \cite{RefITUTerr}, general urban geometry can be effectively described using a standardized statistical model based on three key parameters: the ratio of built-up area to total area ($\alpha_o$), the average number of building density measured in buildings per square kilometer ($\beta_o$), and a building height scale parameter ($\gamma_o$). The latter defines the building height distribution (${h_\mathrm{b}}$) according to the Rayleigh probability density function:

\begin{equation}
P(h_\mathrm{b}) = \frac{h_\mathrm{b}}{\gamma_0}\exp \left(\frac{ - h_\mathrm{b}^2}{2\gamma_0^2}\right),
\label{EQ:1}
\end{equation}

To accurately capture the geometric complexities of 6G deployment scenarios, we selected three distinct environments adapted from \cite{RefITUTerr}: Suburban, Urban, and HighRise Urban. These categories are essential for analyzing the impact of high building densities and vertical structures.

The model construction involves a two-step process:
\begin{itemize}
    \item \textbf{Vertical Geometry:} Building heights are generated randomly according to the statistical distribution defined in (1).
    \item \textbf{Horizontal Layout:} The spatial arrangement is deterministic, calculated using ITU statistical parameters $\alpha_0$ and $\beta_0$.
\end{itemize}

\begin{figure}[t]
    \centering
    \subfloat[]{%
        \includegraphics[width=0.57\linewidth,clip,trim=5 2 -50 0]{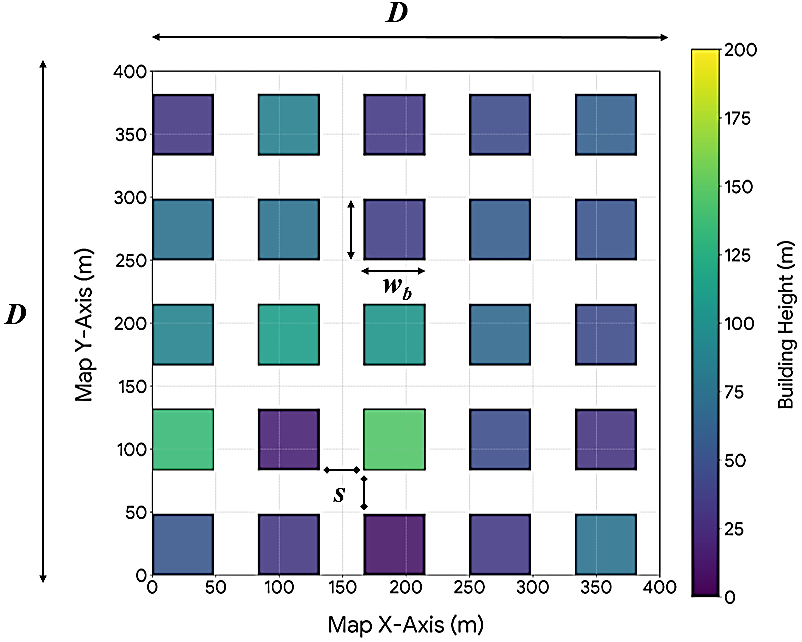}
    }
    \hfill
    \subfloat[]{%
        \includegraphics[width=0.38\linewidth,clip,trim=0 -40 0 0]{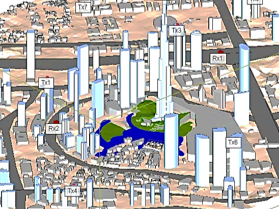}
    }
    \caption{Generated city model based on (a) the ITU statistical model, and (b) 3D CAD model of the Dubai downtown area.}
    \label{fig:Math}
\end{figure}

The specific city-layout parameters such as the building width ($w_b$) in meters, the street width ($s$) in meters, the square network area ($D$) in kilometers, and the number of buildings ($N_b$) are linked to the ITU statistical parameters as follows: $\alpha_0 = \frac{w_b^2 \cdot N_b}{(1000D)^2}$ and $\beta_0 = \frac{N_b}{D^2}$. Solving for the physical dimensions yields $w_b = 1000 \sqrt{\frac{\alpha_0}{\beta_0}}$, $D = \frac{s + w_b}{1000} \sqrt{N_b}$, and $s = \frac{1000}{\sqrt{\beta_0}} - w_b$. Fig. \ref{fig:Math}(a-b) illustrates an example of this generated area, displaying the array of structures and corresponding parameters within a limited region. Using this approach, the realization of a hypothetical city layout is sampled from the ITU statistical model and then imported to the Ray Tracer (RT). The ITU statistical parameters for these environments are summarized in Table \ref{tab:table1}, corresponding to a total network area of $D \times D = 1.2$ km$^2$ containing approximately $N_b$ buildings.
To validate the proposed model, we utilized a 3D CAD representation of downtown Dubai, serving as a characteristic HighRise Urban environment. This model was developed using the Blender-OpenStreetMap (OSM) plugin, which incorporates precise urban geometries alongside realistic material properties and physical features, as shown in Fig.\ref{fig:Math}.

To evaluate both C-V2B, C-P2B performance subject to static building blockages, we analyzed the system under two distinct interference scenarios as follows:
\begin{itemize}
\item Scenario 1 ({\it interference-free}): user equipments (UEs) are allocated orthogonal resources, effectively eliminating inter-cell interference. The received signal quality depends only on thermal noise and propagation characteristics.
\item Scenario 2 ({\it full-Interference}): This represents a worst-case interference profile where all BSs operate on the same frequency resources. Consequently, each UE experiences cumulative interference from all non-serving BSs.
\end{itemize}

\section{CHANNEL MODELING METHODOLOGY}

\begin{table}[t]
\caption{City-Layout Parameters for a Square Network Area of $1.2\,\mathrm{km}^2$ ($D\times D$)}
\label{tab:table1}
\centering
\resizebox{\columnwidth}{!}{%
\begin{tabular}{|l|ccc|ccc|}
\hline
\rowcolor{lightgray}
\textbf{City layout} &
\multicolumn{3}{c|}{\textbf{ITU statistical parameters}} &
\multicolumn{3}{c|}{\textbf{City-layout parameters}} \\ \cline{2-7}
\rowcolor[HTML]{EFEFEF}
&
$\boldsymbol{\alpha_0}$ &
$\boldsymbol{\beta_0}$ &
$\boldsymbol{\gamma_0}$ &
$\boldsymbol{N_b}$ &
$\boldsymbol{w_b}$ (m) &
$\boldsymbol{s}$ (m) \\ \hline
\textbf{HighRise Urban} & 0.5 & 300 & 50 & 432 & 40.82 & 16.91 \\ \hline
\textbf{Urban} & 0.3 & 500 & 15 & 720 & 24.49 & 20.23 \\ \hline
\textbf{Suburban} & 0.1 & 750 & 8 & 1080 & 11.54 & 24.97 \\ \hline
\end{tabular}%
}
\end{table}

The system model utilizes a MIMO downlink configuration employing $N_\mathrm{t}$ transmit antennas per sector at each BS and $N_\mathrm{r}$ receive antennas at each UE. To facilitate advanced spatial processing, a Uniform Rectangular Array (URA) is deployed at the BS, while a Uniform Linear Array (ULA) is integrated into the receiver. MIMO channel models are categorized as either narrowband or wideband, depending on the operational bandwidth relative to the propagation environment. Narrowband models assume a frequency-flat response, whereas wideband models are required to account for the frequency-selective fading characteristic of multipath environments. The choice between these channel modeling is dictated by the channel coherence bandwidth ($B_\mathrm{c}$) which is inversely proportional to the Root-Mean-Square (RMS) delay spread ($\tau_{\text{rms}}$). In diverse urban settings, $\tau_{\text{rms}}$ values typically range from 74.5 ns to 490 ns within the 2.1 to 28.5 GHz frequency range, as documented in ITU-R P.1411-10 (Table 11)~\cite{ITURef}. Defining $B_\mathrm{c}$ as the bandwidth over which the frequency correlation remains above 0.9, the relationship is approximated as $B_\mathrm{c} \approx 1/(5\tau_{\text{rms}})$. Consequently, $B_\mathrm{c}$ for the considered frequency range varies from a few tens of kHz to several tens of MHz. In contrast, practical 3GPP specifications for these frequencies define bandwidths ranging from 20 MHz to 400 MHz. Because these operational bandwidths significantly exceed the calculated $B_\mathrm{c}$, the channel exhibits frequency-selective behavior, necessitating the adoption of a wideband MIMO channel model. The corresponding baseband input-output relationship for a wideband MIMO system is expressed as follows:

\begin{equation}
\mathbf{y}(t) = \int \mathbf{H}(\tau)\,\mathbf{x}(t-\tau)\,d\tau + \mathbf{n}(t),
\end{equation}

where $\mathbf{x}(t)$ denotes the transmitted signal, $\mathbf{y}(t)$ is the received signal, $\mathbf{n}(t)$ represents additive white Gaussian noise (AWGN), and $\mathbf{H}(\tau)$ is the $N_\mathrm{r} \times N_\mathrm{t}$ wideband MIMO channel impulse response matrix. The channel matrix $\mathbf{H}(\tau)$, which characterizes the complex channel gains between the transmit and receive antenna elements, is expressed as:

\begin{equation}
\mathbf{H}(\tau) =
\begin{bmatrix}
h_{1,1}(\tau) & \cdots & h_{1,N_\mathrm{t}}(\tau) \\
\vdots & \ddots & \vdots \\
h_{N_\mathrm{r},1}(\tau) & \cdots & h_{N_\mathrm{r},N_\mathrm{t}}(\tau)
\end{bmatrix}_{N_\mathrm{r} \times N_\mathrm{t}},
\end{equation}

where each channel coefficient $h_{mn}(\tau)$ models the link from the $n$-th transmit antenna to the $m$-th receive antenna and is given by:

\begin{equation}
h_{m,n}(\tau) = \sum_{i=1}^{N_\mathrm{p}} A^{(i)}_{m,n}\,
e^{j \psi^{(i)}_{m,n}}\, \delta\!\big(\tau - \tau^{(i)}_{m,n}\big),
\end{equation}

where $\delta(\cdot)$ denotes the Dirac delta function, $N_\mathrm{p}$ is the number of multipath components between transmit antenna $n$ and receive antenna $m$, and $A^{(i)}_{m,n}$, $\psi^{(i)}_{m,n}$, and $\tau^{(i)}_{m,n}$ represent the amplitude, phase, and delay of the $i$-th path on the $m$--$n$ link, respectively.

\section{Statistical Evaluation of Downlink C-V2B and C-P2B Performance}

This section provides statistical assessments of downlink C-V2B and C-P2B performance within three different environments. The evaluation considers both {\it interference-free} and {\it full-interference} scenarios while accounting for the deterministic impact of static building blockages. Our analysis focuses on the Cumulative Distribution Functions (CDFs) of data rate driven by the SNR and SINR across the investigated frequency bands. To optimize the single-stream MIMO link, Maximum Ratio Transmission (MRT) is adopted for transmit beamforming, while Maximal Ratio Combining (MRC) is utilized at the receiver to maximize spatial diversity. Each element of the beamforming vector is normalized such that the sum of its squared magnitudes equals the number of receive antennas $N_\mathrm{r}$, ensuring a consistent and fair comparison across various combining schemes. The following formulations detail the methodology used to leverage the high-fidelity insights of RT within a wideband MIMO framework. Under the {\it full-interference} scenario, which accounts for simultaneous transmissions from all neighboring BSs, the system performance is quantified by the SINR, expressed as \cite{Ref6}:

\begin{equation}
\mathrm{SINR} = \frac{P_{\mathrm{t}}}{N_\mathrm{t}\,(\sigma_n^2+P_{\text{I,avg}})}\,
\sum_{m=1}^{N_r}\sum_{n=1}^{N_t} \bigl| \mathbf{H}_{m,n} \bigr|^2,
\label{eq:SINR}
\end{equation}

\noindent where $P_{\mathrm{t}}$ is the total transmit power, $\sigma_n^2 = N_0 B$ denotes the noise variance, $N_0$ is thermal noise spectral density ($10^{-21} \rm{A}^{2}/\rm{Hz}$), $B$ is the channel bandwidth, and $P_\mathrm{I,avg}$ represents the average interference power from $N_\mathrm{I}$ interfering BSs, given by

\begin{equation}
P_{\text{I,avg}} =
\sum_{j=1}^{N_{\mathrm{I}}} \frac{P_{\mathrm{t},j}}{N_{\mathrm{t},j}}
\left[ \frac{1}{N_\mathrm{r}} \sum_{u=1}^{N_{\mathrm{t},j}} \sum_{m=1}^{N_\mathrm{r}} |\mathbf{H}_{j,u,m}|^2 \right],
\end{equation}

\noindent with $P_{\mathrm{t},j}$ denoting the transmit power of the $j$-th interfering BS, $N_{\mathrm{t},j}$ the number of elements of the $j$-th interfering BS, and $\mathbf{H}_{j,u,m}$ the channel coefficient between the $u$-th element of the $j$-th BS and the $m$-th element of the UE. Note that in the absence of average interference power, i.e., when $P_\mathrm{I,avg}=0$, the SINR simplifies to SNR. Based on these KPIs, the achievable data rate $R$ is computed according to the Shannon capacity \cite{Refrate}:

\begin{equation}
R = B \min \big(\rho_{\mathrm{max}}, \; \alpha \log_2(1+\Gamma) \big),
\label{eq:Rate}
\end{equation}

where $\alpha$ represents a bandwidth loss factor that accounts for protocol overhead and receiver imperfections, $\rho_{\max}$ is the maximum achievable spectral efficiency, and $\Gamma$ represents the SNR or SINR, corresponding to {\it interference-free} or {\it full-interference} scenarios, respectively.

\begin{figure}[t]
\centering
\includegraphics[clip,trim=7 15 0 5,scale=0.245]{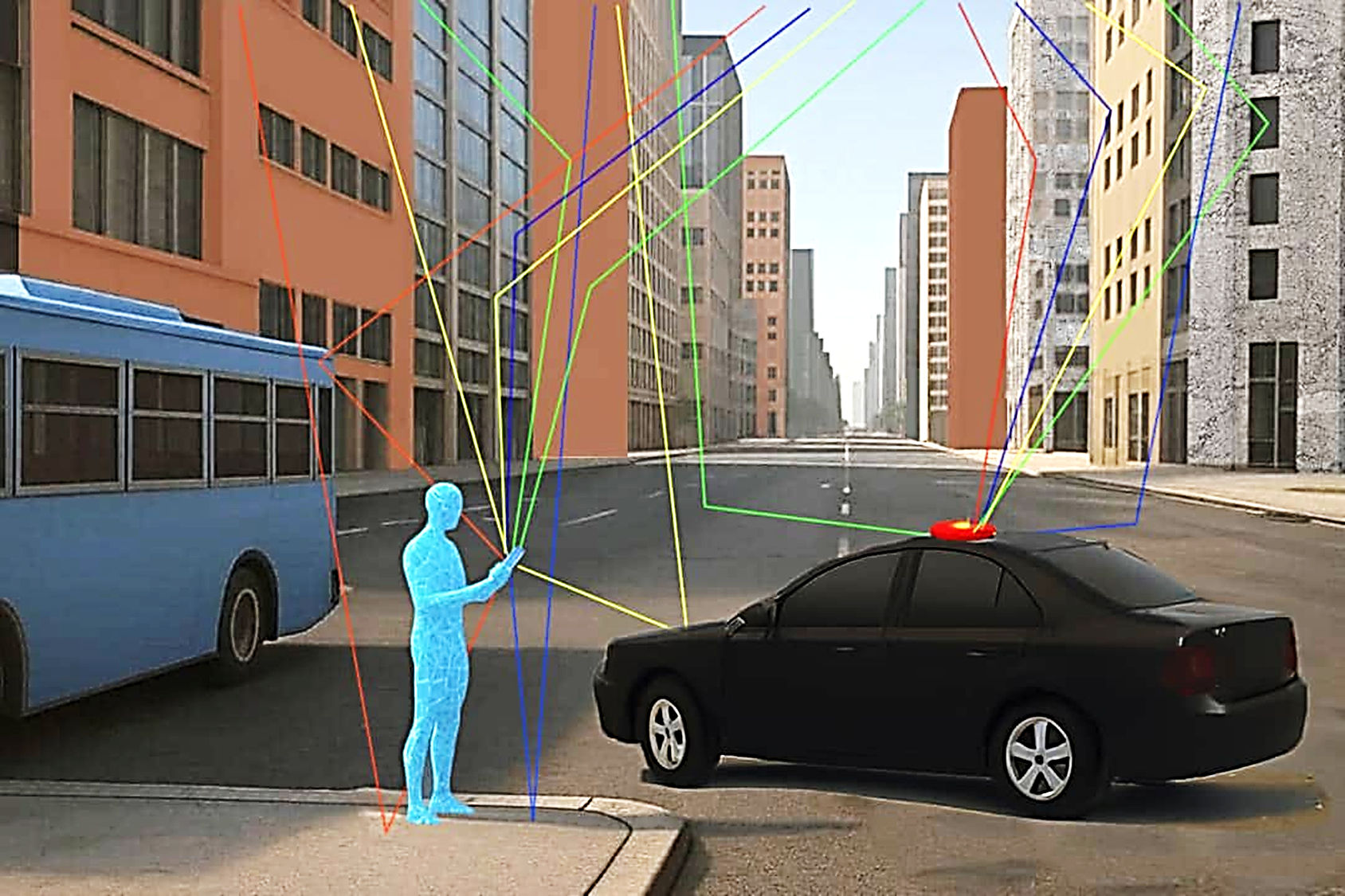}
\caption{The vehicle and human CAD models along with their associated antenna configurations used in the Wireless InSite RT simulator for urban network scenarios.}
\label{fig:UE-CADs2}
\end{figure}

\begin{figure}[t]
\centering
\subfloat[]{%
\includegraphics[width=0.96\columnwidth,clip,trim=10 1 6 3]{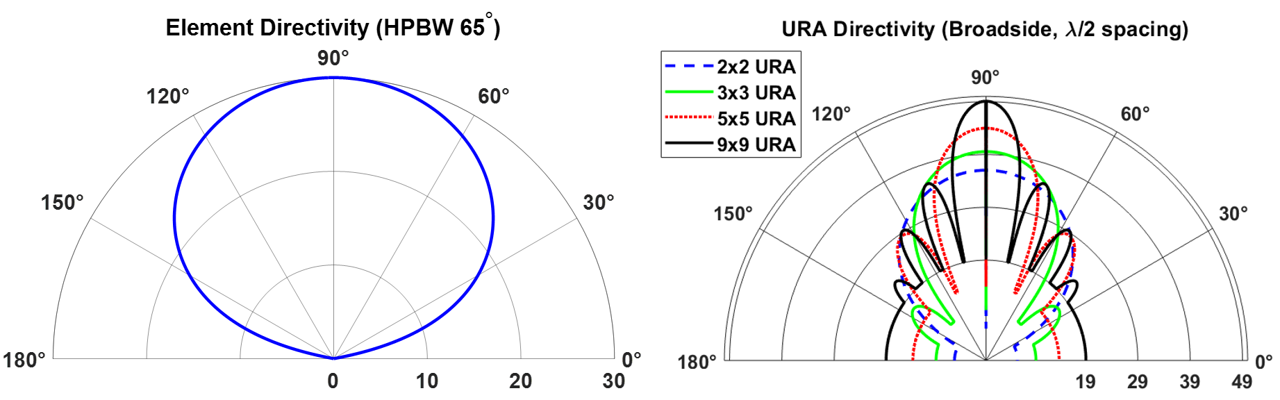}
}

\subfloat[]{%
\includegraphics[width=0.96\columnwidth,clip,trim=-20 1 6 8]{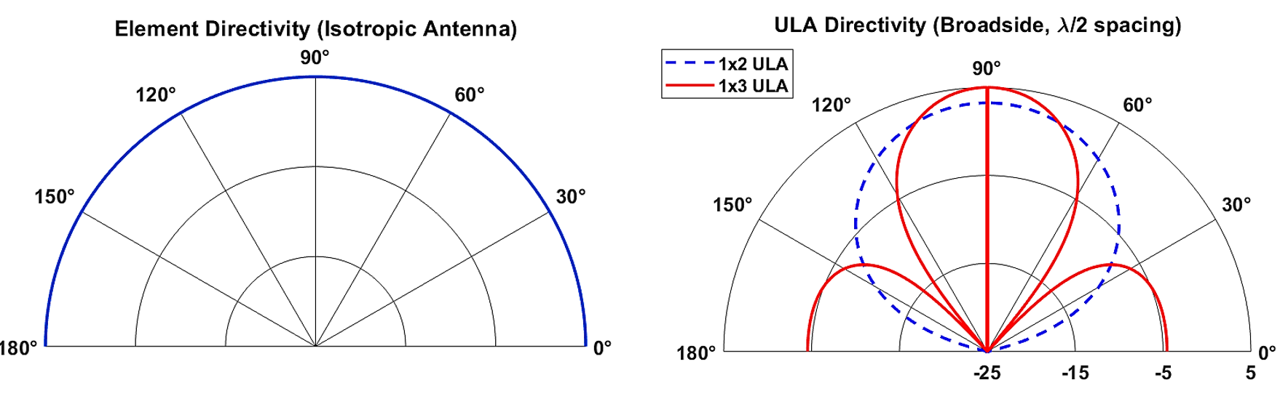}
}
\caption{Directivity patterns for single antenna element and antenna arrays: (a) BS configurations with 2$\times$2, 3$\times$3, 5$\times$5, and 9$\times$9 URA, and (b) Vehicle-UE configurations with 1$\times$2 and 1$\times$3 ULA.}
\label{fig:AntPat}
\end{figure}

In all simulations the serving BS for each UE is assigned as the one providing the highest signal quality. The evaluations are conducted using both ITU statistical parameters, and the 3D CAD model of the Dubai downtown area corresponding to different urban environments (including HighRise Urban, Urban, Suburban) and a realistic HighRise Urban layout illustrated in Fig. \ref{fig:Math}(b-c), respectively. RT simulations are conducted via Wireless Insite with the engine configured to capture a maximum of 25 paths per link. The simulator accounts for 6 reflections, 1 diffraction, and 1 transmission per path with a minimum power threshold of~-250 dBm. Table \ref{tab:table2} lists the conductivity (S/m) and relative permittivity (denoted as $\sigma$ and $\varepsilon_r$) of various materials including concrete, glass, wood, brick, and dry earth ground across different frequencies \cite{Ref10}. The network extends a 1.2$\times$1.2 km\textsuperscript{2} area, featuring 17 BSs deployed in a hexagonal grid with an Inter-Site Distance (ISD) of 350 m \cite{Ref3}. The rooftop-mounted antennas have a Half-Power BeamWidth (HPBW) of 65$^\circ$ and a maximum element gain of 30 dBi with a downtilt of --12$^\circ$. To maintain a consistent transmitter aperture size for fair performance comparison across frequencies, the number of antenna elements is scaled accordingly (see Table \ref{tab:table3}). Furthermore, the allocated bandwidth is increased in proportion to the carrier frequency, reflecting standardized spectrum allocation policies. Fig. \ref{fig:UE-CADs2} illustrates the CAD models of the vehicle and pedestrian in an urban network scenario. The vehicle UE is equipped with a roof-mounted isotropic antenna, while the pedestrian UE follows the one-hand-grip configuration specified in 3GPP TR 38.901 \cite{Refgriphand}, employing a directional antenna with an HPBW of 125$^\circ$ and a maximum gain of 5.3 dBi. Fig. \ref{fig:AntPat} shows the element-wise and array-wise directivity patterns for the BS and UE respectively, where the aperture directivity is presented for various configurations. To ensure the accuracy of the RT simulations, 370 UEs are randomly scattered across the network area following a uniform distribution, where the results are obtained using 10 deployments of the city layout. Table \ref{tab:table3} summarizes the simulation parameters for C-V2B downlink system operating across different frequencies.

\begin{table}[t]
\caption{Material properties at different frequencies}
\label{tab:table2}
\centering
\resizebox{\columnwidth}{!}{%
\begin{tabular}{|l|cc|cc|cc|cc|}
\hline
\rowcolor[HTML]{EFEFEF}
\textbf{Material} &
\multicolumn{2}{c|}{\textbf{4.6 GHz}} &
\multicolumn{2}{c|}{\textbf{8.2 GHz}} &
\multicolumn{2}{c|}{\textbf{15 GHz}} &
\multicolumn{2}{c|}{\textbf{28 GHz}} \\ \cline{2-9}
\rowcolor[HTML]{EFEFEF}
& \textbf{$\sigma$ (S/m)} & \textbf{$\varepsilon_r$}
& \textbf{$\sigma$ (S/m)} & \textbf{$\varepsilon_r$}
& \textbf{$\sigma$ (S/m)} & \textbf{$\varepsilon_r$}
& \textbf{$\sigma$ (S/m)} & \textbf{$\varepsilon_r$} \\ \hline
Concrete   & 0.14  & 5.24 & 0.23  & 5.24 & 0.38  & 5.24 & 0.63  & 5.24 \\
Glass      & 0.03  & 6.31 & 0.06  & 6.31 & 0.12  & 6.31 & 0.24  & 6.31 \\
Brick      & 0.03  & 3.91 & 0.03  & 3.91 & 0.04  & 3.91 & 0.04  & 3.91 \\
Dry earth  & 0.003 & 3.00 & 0.01  & 3.00 & 0.036 & 3.00 & 0.147 & 3.00 \\ \hline
\end{tabular}%
}
\end{table}

\begin{table}[t]
\caption{Simulation Parameters for MIMO C\,-V2B and C-P2B Communications}
\label{tab:table3}
\resizebox{\columnwidth}{!}{
\begin{tabular}{|c|c|c|c|c|}
\hline
\textbf{Parameter} & \multicolumn{4}{c|}{\textbf{Value / Description}} \\ \hline
\multicolumn{5}{|c|}{\cellcolor[HTML]{EFEFEF}\textbf{Frequency and Environment}} \\ \hline
Network area (km $\times$ km) & \multicolumn{4}{c|}{$1.2 \times 1.2$ km$^2$} \\ \hline
ISD (m) & \multicolumn{4}{c|}{350 m} \\ \hline
Number of cell sectors & \multicolumn{4}{c|}{3} \\ \hline
Power spectral density of noise (A$^2$/Hz) & \multicolumn{4}{c|}{$10^{-21}$ A$^2$/Hz} \\ \hline
Frequency bands (GHz) & 4.6 GHz & 8.2 GHz & 15 GHz & 28 GHz \\ \hline
Bandwidth (MHz) & 60 MHz & 200 MHz & 300 MHz & 400 MHz \\ \hline
\multicolumn{5}{|c|}{\cellcolor[HTML]{EFEFEF}\textbf{Base Station}} \\ \hline
Location & \multicolumn{4}{c|}{Rooftop} \\ \hline
Antenna element radiation pattern & \multicolumn{4}{c|}{\begin{tabular}[c]{@{}c@{}}HPBW $65^\circ$, maximum element gain 30 dBi\\ (3GPP TR 37.840)\end{tabular}} \\ \hline
Antenna tilt angle & \multicolumn{4}{c|}{$-12^\circ$} \\ \hline
\multirow{2}{*}{URA dimensions} & 4.6 GHz & 8.2 GHz & 15 GHz & 28 GHz \\ \cline{2-5}
& $2\times2$ & $3\times3$ & $5\times5$ & $9\times9$ \\ \hline
\multicolumn{5}{|c|}{\cellcolor[HTML]{EFEFEF}\textbf{User Equipment}} \\ \hline
Location & \multicolumn{4}{c|}{Random locations following a uniform distribution} \\ \hline
Height (m) & \multicolumn{4}{c|}{2 m (3GPP TR 37.840)} \\ \hline
Antenna element radiation pattern & \multicolumn{4}{c|}{\begin{tabular}[c]{@{}c@{}}C-V2B: Isotropic antenna (3GPP TR 37.840)\\ C-P2B: Directional antenna, HPBW $65^\circ$, maximum element \\ gain 30 dBi (3GPP TR 38.901)\end{tabular}} \\ \hline
\multirow{2}{*}{ULA dimensions} & 4.6 GHz & 8.2 GHz & 15 GHz & 28 GHz \\ \cline{2-5}
& \multicolumn{2}{c|}{$1\times2$} & \multicolumn{2}{c|}{$1\times3$} \\ \hline
\end{tabular}
}
\end{table}

\begin{figure*}[t]
\centering
\subfloat[]{\includegraphics[width=.228\textwidth,clip,trim=5 0 9 5]{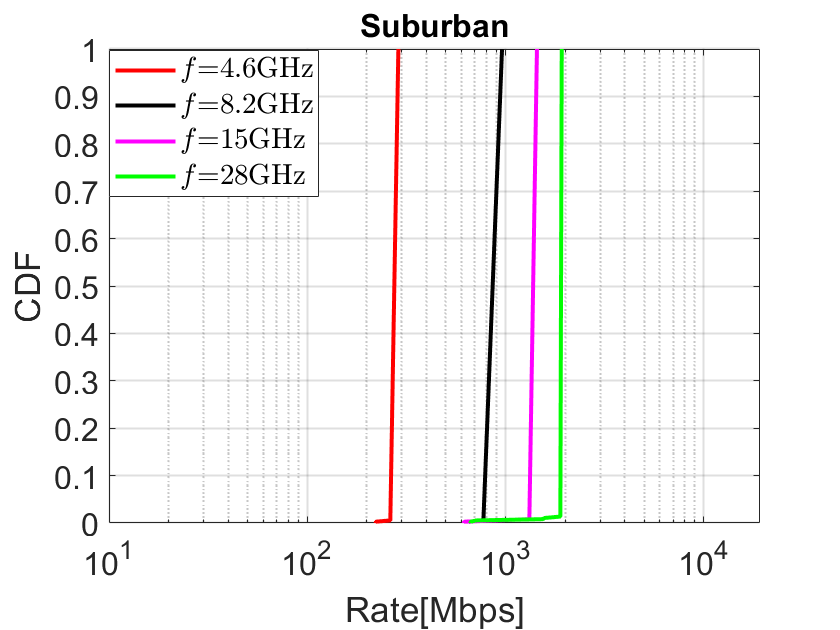}\label{subfig:5a}}\hfill
\subfloat[]{\includegraphics[width=.228\textwidth,clip,trim=10 0 9 1]{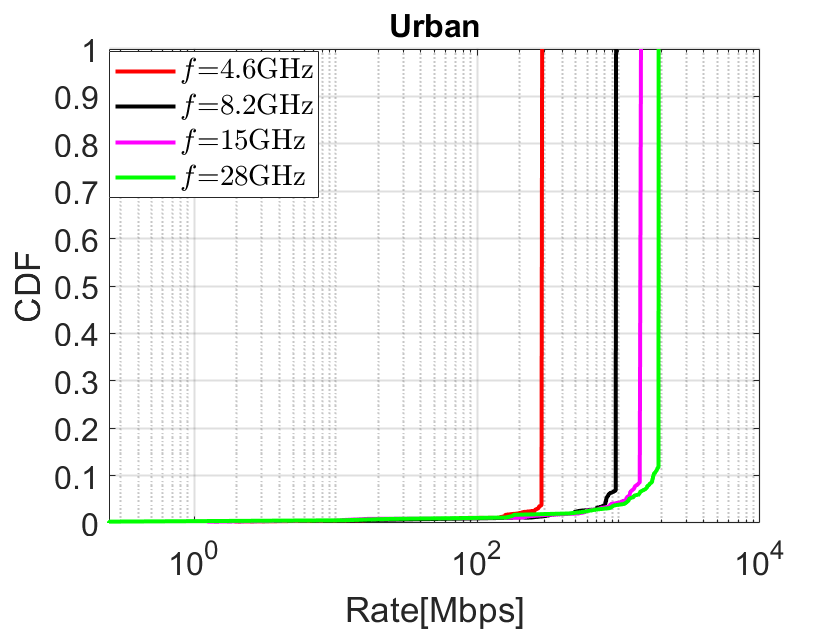}\label{subfig:5b}}\hfill
\subfloat[]{\includegraphics[width=.228\textwidth,clip,trim=8 1 9 5]{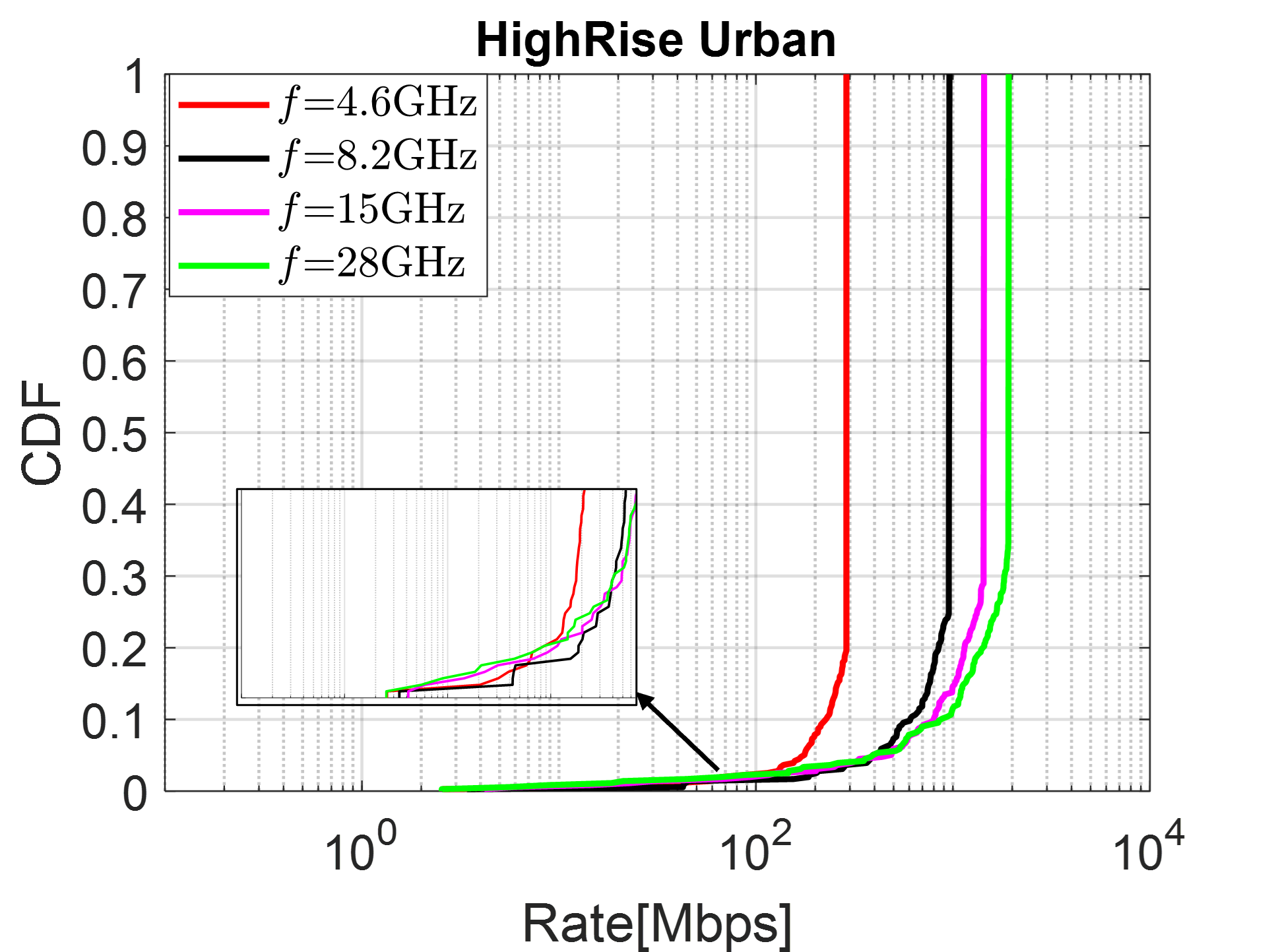}\label{subfig:5c}}\hfill
\subfloat[]{\includegraphics[width=.228\textwidth,clip,trim=2 1 9 5]{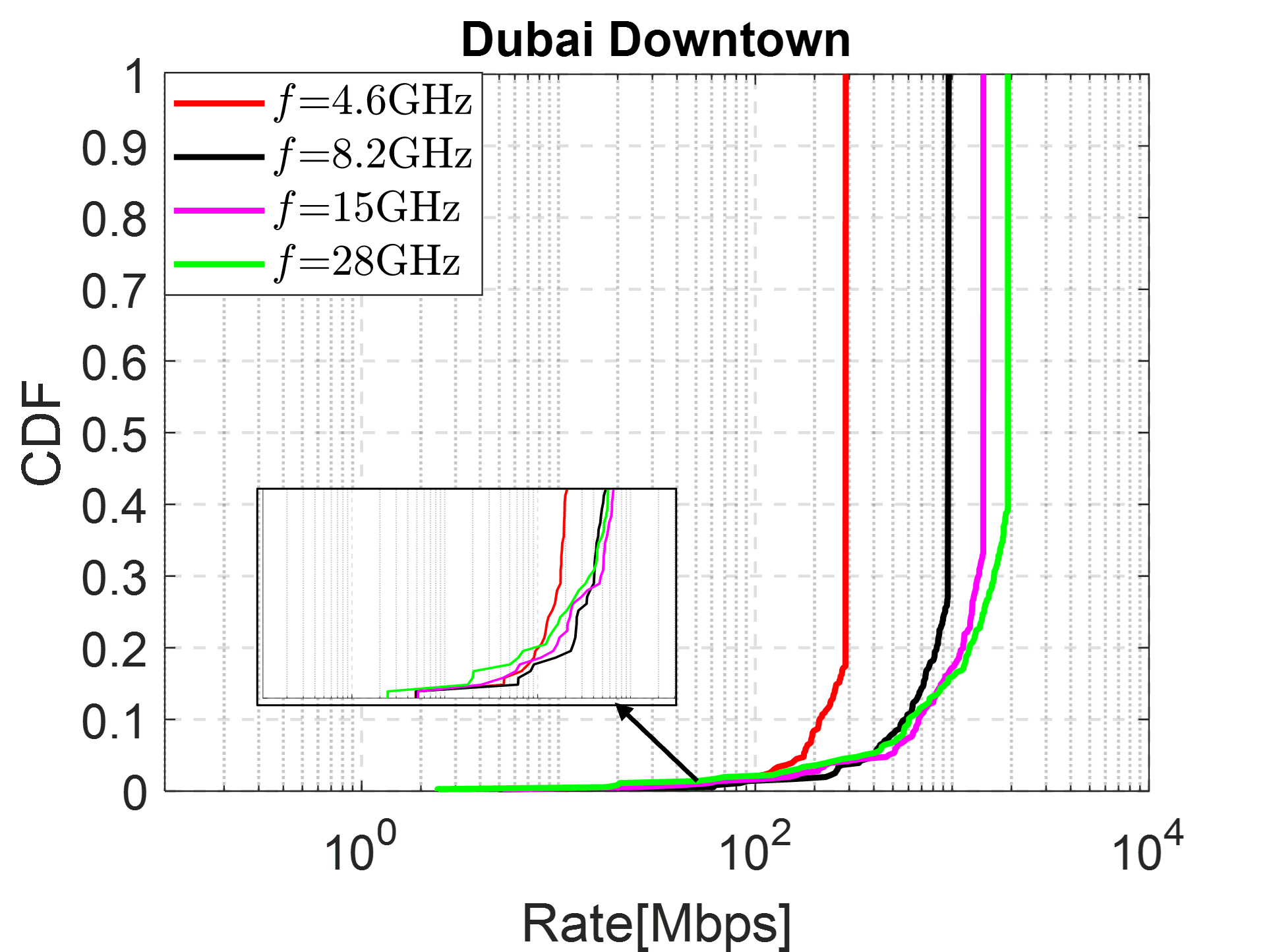}\label{subfig:5d}}

\subfloat[]{\includegraphics[width=.228\textwidth,clip,trim=10 0 9 5]{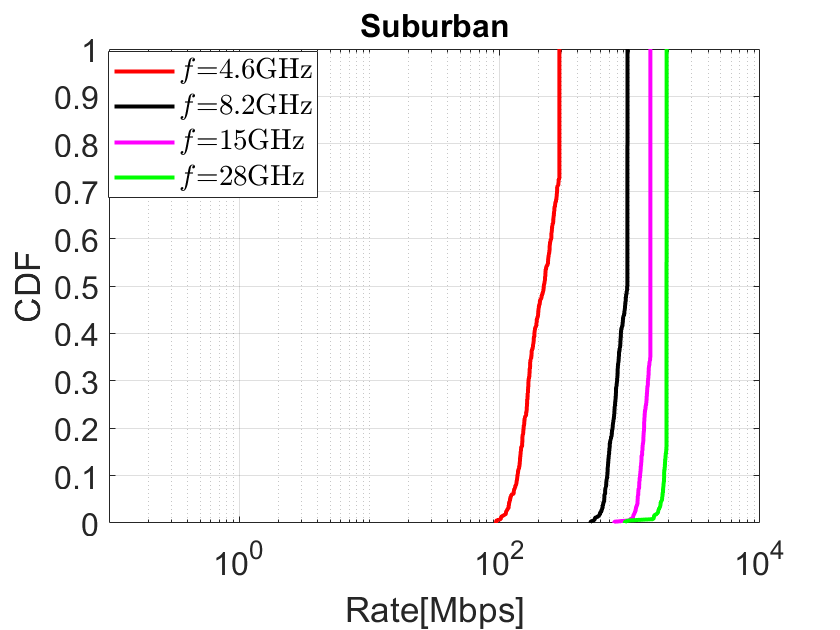}\label{subfig:5e}}\hfill
\subfloat[]{\includegraphics[width=.228\textwidth,clip,trim=10 0 9 5]{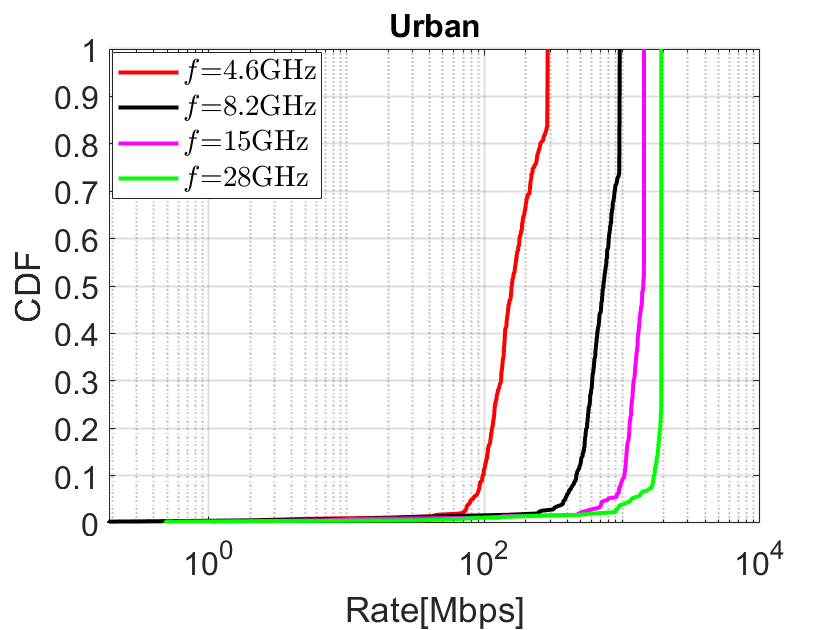}\label{subfig:5f}}\hfill
\subfloat[]{\includegraphics[width=.228\textwidth,clip,trim=0 1 9 5]{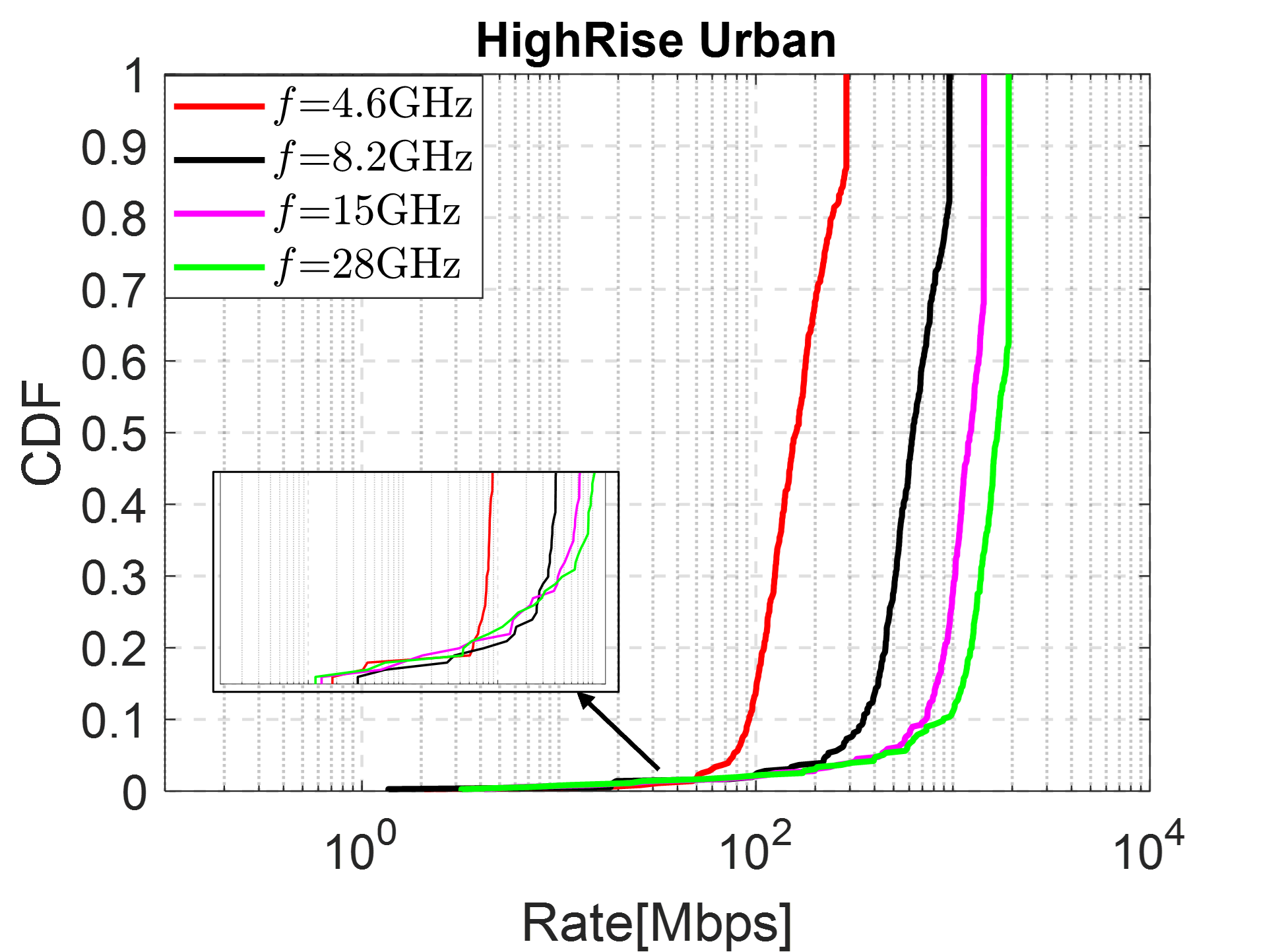}\label{subfig:5g}}\hfill
\subfloat[]{\includegraphics[width=.228\textwidth,clip,trim=5 1 9 5]{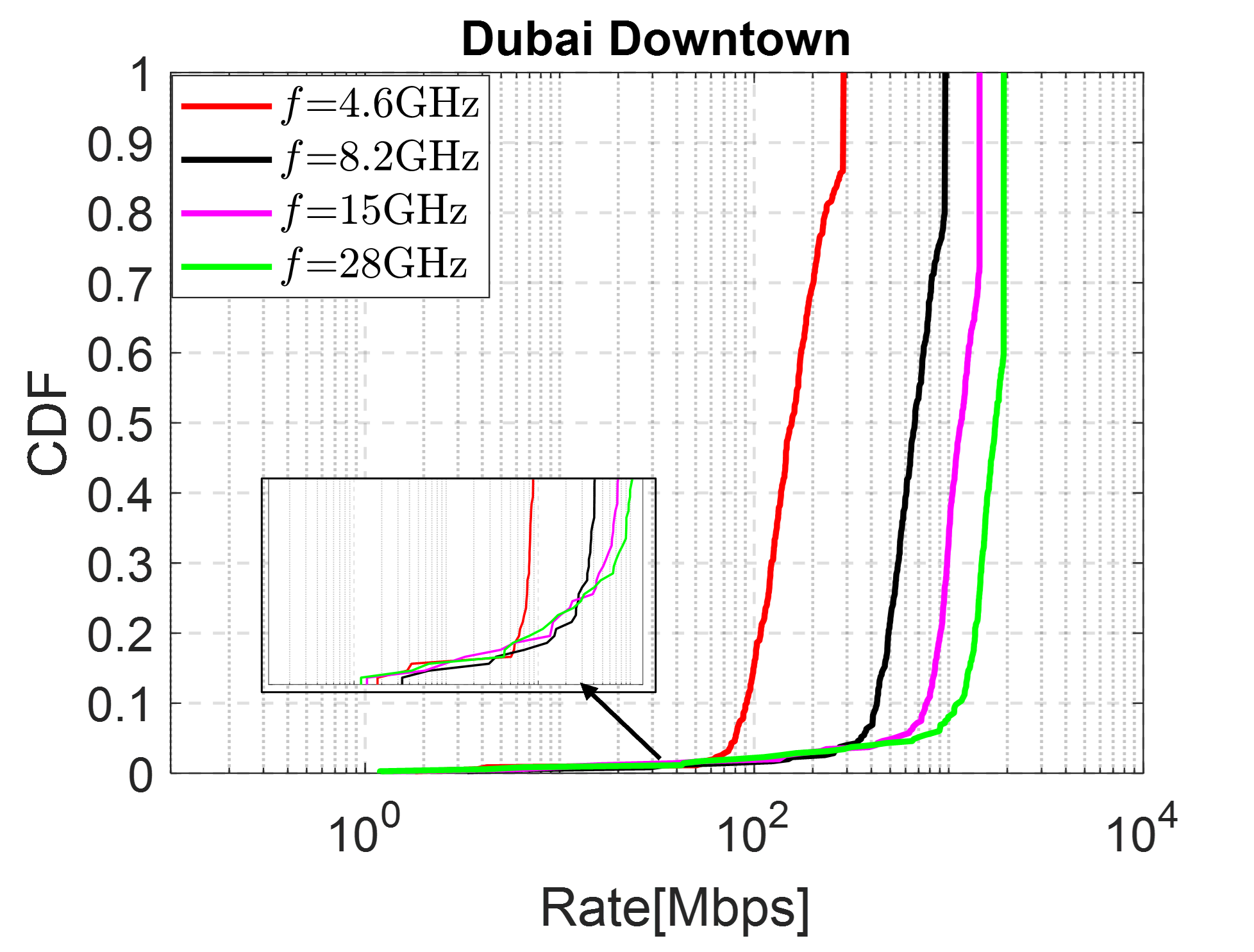}\label{subfig:5h}}
\caption{(a--d) CDFs of data rate under the interference-free scenario; (e--h) CDFs of data rate under the full-interference scenario across Suburban, Urban, and HighRise Urban environments, considering both the ITU statistical model and the 3D CAD model of Dubai downtown at different frequencies.}
\label{fig:RATE}
\end{figure*}

\subsection{FR3 vs. Non-FR3 Band Performance for downlink C-V2B Networks}

We compute the wideband SNR and SINR for each UE--BS pair under {\it interference-free} and {\it full-interference} scenarios considering static blockage and assume that each UE is served by the BS offering the strongest quality. Fig. \ref{fig:RATE} shows the CDFs of user data rates cross all UE locations in three distinct scenarios: Suburban, Urban, and HighRise Urban across the FR1 (sub-6 GHz), FR3, and FR2 (mmWave) bands at 4.6, 8.2~-~15, and 28 GHz, respectively. Figs. \ref{fig:RATE}(a--d) present the CDFs of the data rate $R$ for the {\it interference-free} scenario, while Figs. \ref{fig:RATE}(e--h) illustrate the CDFs of $R$ under the {\it full-interference} scenario, evaluated using the ITU statistical model and the 3D CAD model of the Dubai downtown area, respectively.

The user data rate formulated in (\ref{eq:Rate}) is based on the realistic system model described in \cite{Refrate}, with parameters $\alpha = 0.57$ and $\rho_{\max} = 4.8$. From (\ref{eq:Rate}), the maximum achievable rate is $R = B \cdot \rho_{\max}$ which scales linearly with bandwidth. Consequently, although UEs operating at higher frequencies experience lower SNRs due to the combined influence of several factors, including path loss in numerator of (\ref{eq:SINR}), antenna gain, and noise power (i.e., $N_{0}B$) in the denominator, their wider bandwidths yield significantly higher data rates. In contrast, achievable rates remain limited despite higher SNR levels at lower frequencies (4.6 GHz) due to narrower bandwidth allocation and wide beamwidth exacerbate inter-cell interference. For example, at the 0.5 CDF level in the HighRise Urban environment (Fig. \ref{fig:RATE}(g-h)), the 15 Ghz and 28 GHz bands achieve data rates exceeding 1 Gbps, whereas 4.6 GHz remains at approximetly 180 Mbps. Moreover, as shown in Fig. \ref{fig:RATE}, the maximum achievable rates across all environments increase from about $0.28$\,Gbps at 4.6\,GHz, to about $0.95$\,Gbps at 8.2\,GHz, $1.42$\,Gbps at 15\,GHz, and nearly $1.91$\,Gbps at 28\,GHz. Furthermore, the performance gap between {\it interference-free} and {\it full-interference} scenarios becomes smaller at higher frequencies, since narrow beams at FR3 and mmWave bands inherently mitigate inter-cell interference.

Regarding environmental complexity, variations in building density and height as static blockages further influence the rate CDFs. Suburban areas consistently achieve the highest and most uniform rates, reaching nearly 1.91 Gbps at 28 GHz for the top 95\% of UEs. This performance is attributed to dominant LOS conditions and minimal static blockage, resulting in most UEs experiencing nearly constant rates under both {\it interference-free} and {\it full-interference} scenarios, see Fig. \ref{fig:RATE}(a and e). Urban environments provide intermediate performance, where moderate reflections maintain connectivity but multipath and partial NLoS effects broaden the CDFs comparing to Suburban areas, as illustrated in Fig. \ref{fig:RATE}(b and f). Notably, across both free/full-interference scenarios in urban environment, UEs at the cell edge (bottom 0.1 CDF level) can approximately achieve the maximum achievable rate.

HighRise Urban exhibits the strongest rate degradation as shown in Fig. \ref{fig:RATE}(c-d) and (g-h), particularly below the 0.1 of CDFs where rates drop to low-hundreds of Mbps at 4.6 GHz due to severe blockage and NLoS conditions. Figs. \ref{fig:RATE}(c-d) reveal a detailed performance trade-off between bandwidth gains and propagation losses across the investigated frequency bands. For the most of UEs, particularly those located within the top 70\% of the distribution, higher frequencies such as 15 GHz and 28 GHz demonstrate superior performance. Although these bands suffer from more pronounced attenuation in {\it interference-free} scenario, their significantly larger bandwidth allocations allow them to achieve data rates exceeding 1 Gbps. Therefore, the data rate scales positively with carrier frequency, allowing the FR3 and mmWave band to comfortably outperform sub-6 GHz baseline across both the ITU statistical model and the 3D CAD model of the Dubai downtown area. However, a distinct frequency-order inversion occurs at the bottom 0.1 CDF level corresponding to cell edge UEs compared to non-cell-edge UEs. For these users, the severe signal attenuation associated with 15 GHz and 28 GHz leads to a low SNR that effectively offsets the advantages of a wider bandwidth. Consequently, at the cell edge, operating at higher frequencies fail to surpass the data rates provided by the 4.6 GHz band. Notably, the 8.2 GHz (FR3) band emerges as a peak performance point for cell-edge UEs, as it achieves higher data rates than both the sub-6 GHz and mmWave alternatives by balancing favorable propagation characteristics with moderate bandwidth improvements. A critical turning point is observed around the 0.1 CDF level, beyond which higher-frequency bands begin to dominate in performance. After this point, the 15 GHz (FR3) band achieves data rates remarkably close to those of the mmWave band. This indicates that operating within the upper FR3 spectrum can offer a performance profile comparable to mmWave systems, while potentially providing improved coverage for both non-cell edge and cell-edge UEs. In the {\it full-interference} scenario shown in Fig. \ref{fig:RATE}(g-h), inter-cell interference causes a substantial reduction in achievable rates compared to {\it interference-free} scenario. This impact is particularly severe at lower frequencies, where wider beamwidths capture more interference, resulting in a performance drop for approximately 55--60\% of UEs across both the ITU statistical and 3D CAD models. Conversely, the degradation is notably less severe at higher frequencies, affecting only about 30--40\% of UEs, owing to the narrower beamwidths enabled by accommodating more antenna elements within the same transmitter aperture size. The performance trends at the cell edge UEs at the bottom 0.1 CDF level align closely with those observed in the {\it interference-free} scenario. The data rates for high-frequency bands remain restricted by severe propagation losses that outweigh their bandwidth advantages and beamforming gain. A more detailed inspection of the results reveals that 2\% of UEs operating in the FR3 band can achieve rates of 150 Mbps; however, at 28 GHz, these rates drop significantly to below 70 Mbps. These findings highlight the dual nature of mmWave operation: while mmWave band enables multi-Gbps peak rates for non-cell-edge UEs with favorable channel conditions, they present a distinct risk of leaving cell-edge UEs under served. This disparity further suggests that operating in the FR3 bands with advanced beam management or relaying techniques ensure consistent service across the entire network footprint.

\begin{figure}[t]
\centering
\includegraphics[clip,trim=0 1 0 8,scale=0.24]{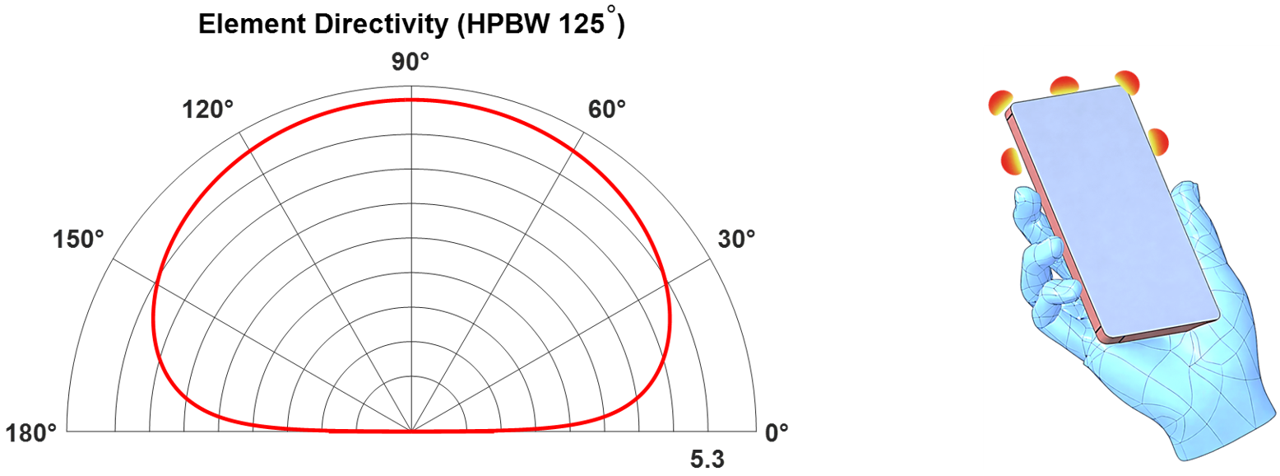}
\caption{One-hand-grip pedestrian-UE antenna radiation pattern with HPBW 125$^\circ$ and 5.3 dBi directional gain (left) with the locations of the five antenna elements on the cellphone (right).}
\label{fig:cellphone}
\end{figure}

\begin{figure*}[t]
\centering
\includegraphics[width=0.94\textwidth,clip,trim=5 5 5 10]{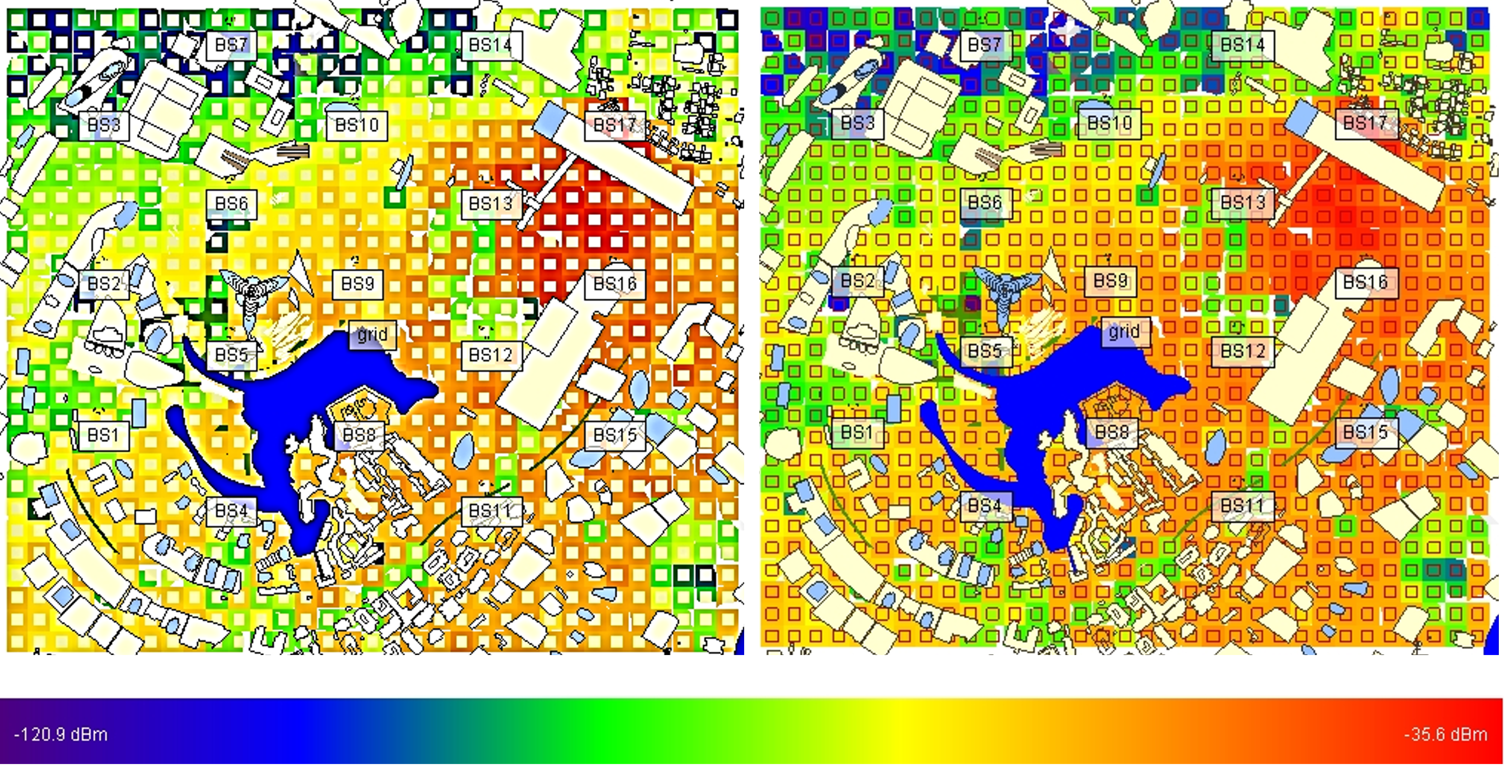}
\caption{Comparison of coverage maps at 8.2 GHz for vehicular (left) and pedestrian (right) UEs across the 3D CAD model of Dubai Downtown.}
\label{fig:heatmap}
\end{figure*}
Based on the results presented in Fig. \ref{fig:RATE}, three key observations can be summarized as follows:
\begin{itemize}
    \item {\bf Observation 1:} {\it In Suburban, Urban, and the upper o.1 CDF level in HighRise Urban environments, the higher carrier frequencies yield higher data rates for non-cell-edge UEs.}
    \item {\bf Observation 2:} {\it At the lower 0.1 CDF level, corresponding to UEs located close to the cell edge in HighRise Urban area, the curves exhibit a frequency-order inversion compared to non-cell-edge UEs. Specifically, the data rate degrades with increasing carrier frequency, indicating a performance reversal relative to Observation 1.}
    \item {\bf Observation 3 (FR3 Analysis):} {\it Both the ITU statistical models for Suburban, Urban, and HighRise Urban environments and the 3D CAD model of Dubai Downtown show that, for non-cell-edge UEs, the data rates achieved in the FR3 band are very close to the best rates obtained in the mmWave bands, despite FR3 benefiting from less bandwidth. In contrast, for cell-edge UEs, operation in the FR3 band yields the highest data rates among all considered bands, outperforming both sub-6 GHz and mmWave.}
\end{itemize}

\subsection{Coverage Probability: Vehicular vs. Pedestrian UEs}

Based on the statistical evaluation of interference explained in Section IV, the coverage probability is defined as the probability that the SINR of UEs exceeds a given threshold $\gamma^{\mathrm{th}}$, expressed as:

\begin{equation}
p^{\mathrm{cov}}(\gamma^{\mathrm{th}}) = \Pr \left[ \mathrm{SINR}^{\mathrm{UE}} > \gamma^{\mathrm{th}} \right],
\end{equation}

where the SINR is calculated by (\ref{eq:SINR}), and $\gamma^{\mathrm{th}}$ is set to 10 dB.

\begin{figure}[t]
\centering
\includegraphics[width=0.90\columnwidth,clip,trim=20 25 45 3]{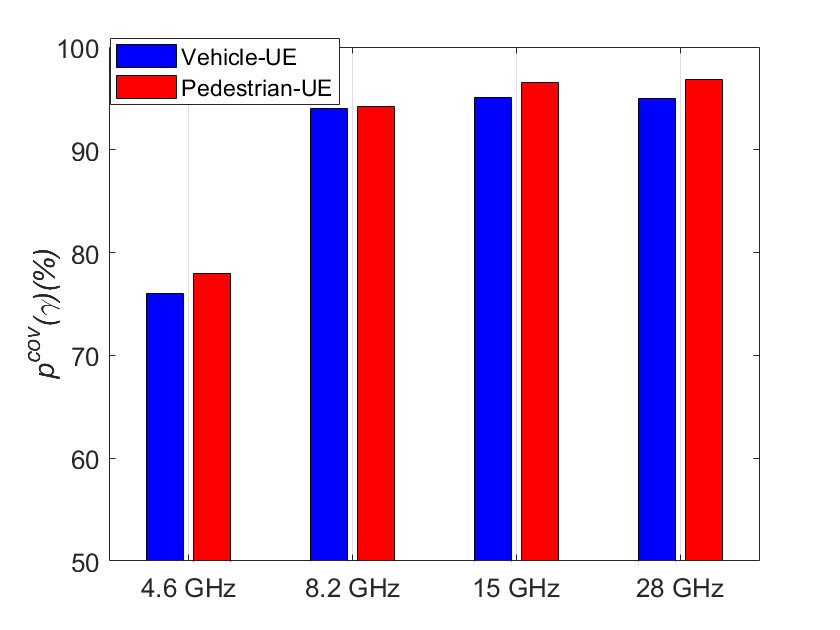}
\caption{Coverage probability comparison between vehicular and pedestrian UEs across different frequency bands including sub-6 GHz, FR3, and mmWave.}
\label{fig:bar}
\end{figure}

Fig. \ref{fig:cellphone} shows the one-hand-grip pedestrian UE antenna model following the configuration specified in 3GPP TR 38.901 \cite{Refgriphand}, exhibiting a wide radiation pattern with a 125$^\circ$ HPBW and 5.3 dBi directional gain. The placement of five antenna elements around the cellphone provides spatial diversity, improving link reliability. The received-power heat maps in Fig. \ref{fig:heatmap} show negligible impact on coverage probability for vehicular and pedestrian UEs at 8.2 GHz, as expected since path loss and static blockage are primarily driven by the environment and the rooftop-mounted BS deployment rather than UE mobility type. However, pedestrian UEs exhibit slightly greater resilience and uniformly distributed received power. This is because they are more likely to remain in LoS regions and are less frequently shadowed by dynamic obstructions compared to vehicular UEs, which experience additional blockage from vehicles and street-level clutter. Overall, changing the UE type does not significantly alter the coverage pattern, but pedestrian UEs benefit from slightly smoother power variations. The results in Fig. \ref{fig:bar} also indicate that UE mobility and height (vehicular vs. pedestrian) have a negligible impact on SINR-based coverage within the macro-cell deployment. Across all frequency bands (4.6, 8.2, 15, and 28 GHz), pedestrian UEs consistently achieve a slightly higher coverage probability about 1\%~-~3\% than vehicular UEs. While the performance gap is more pronounced at FR1, where vehicular UEs are more vulnerable to interference and dynamic shadowing associated with wider beam patterns, operating at FR3 provides the minimum differences.

\begin{itemize}
    \item {\bf Observation:} {\it Incorporating the one-hand-grip pedestrian UE model into the RT tool indicates that the transition from vehicular to pedestrian UEs within a macro-cell deployment leads to negligible differences in coverage probability with the minimum variation observed in FR3, particularly at 8.2 GHz.}
\end{itemize}

\section{CONCLUSION}

We performed a detailed FR3 channel characterization in different urban environments and benchmarked its performance against FR1 and FR2 bands for C-V2B and C-P2B downlink communications. Using Wireless InSite ray-tracing simulations with high-resolution 3D CAD models and MIMO beamforming, UE data rate and coverage probability were analyzed under multiple interference conditions. The results suggest that achieving optimal performance across diverse scenarios may necessitate dynamic switching between low- and high-frequency bands, which in turn increases system cost, spectrum licensing requirements, and operational complexity. As an alternative, FR3 provides a well-balanced compromise. Under equal antenna aperture constraints, although mmWave benefits from higher array gains due to a larger number of antenna elements, FR3 surpasses mmWave in both interference-free and full-interference scenarios at low CDF values (data rate CDF $<$ 0.1), corresponding to cell-edge users. These findings indicate that the additional array gain at mmWave frequencies is generally insufficient to compensate for its pronounced path loss in practical deployment scenarios.

\bibliographystyle{IEEEtran}
\bibliography{Rio}

\end{document}